\documentclass[twocolumn]{aastex61}
\usepackage{longtable}
\usepackage{float}
\usepackage{tablefootnote}
\usepackage{enumitem}

\usepackage{hyperref}
\usepackage{bookmark}
\usepackage{amsmath}

\DeclareUnicodeCharacter{2212}{-} 

\submitjournal{ApJ}

\shorttitle{An Analysis of the Radius Gap in planet-hosting M dwarf stars}
\shortauthors{Wanderley et al.}

\begin{document}

\title{An Analysis of the Radius Gap in a sample of Kepler, K2 and TESS Exoplanets Orbiting M dwarf stars}

\correspondingauthor{Fábio Wanderley}
\email{fabiowanderley@on.br}

\author[0000-0003-0697-2209]{Fábio Wanderley}
\affiliation{Observatório Nacional/MCTIC, R. Gen. José Cristino, 77, 20921-400, Rio de Janeiro, Brazil}

\author[0000-0001-6476-0576]{Katia Cunha}
\affiliation{Steward Observatory, University of Arizona, 933 North Cherry Avenue, Tucson, AZ 85721-0065, USA}
\affiliation{Observatório Nacional/MCTIC, R. Gen. José Cristino, 77,  20921-400, Rio de Janeiro, Brazil}

\author[0000-0002-0134-2024]{Verne V. Smith}
\affiliation{NSF’s NOIRLab, 950 N. Cherry Ave. Tucson, AZ 85719 USA}

\author[0000-0002-7883-5425]{Diogo Souto}
\affiliation{Departamento de F\'isica, Universidade Federal de Sergipe, Av. Marcelo Deda Chagas, S/N Cep 49.107-230, S\~ao Crist\'ov\~ao, SE, Brazil}

\author[0000-0001-7962-1683]{I. Pascucci}
\affiliation{Lunar \& Planetary Lab, University of Arizona, Tucson, AZ 85721 USA}

\author[0000-0003-0012-9093]{Aida Behmard}
\affiliation{Center for Computational Astrophysics, Flatiron Institute, 162 Fifth Ave., New York, NY 10010, USA}
\affiliation{American Museum of Natural History, 200 Central Park West, Manhattan, NY 10024, USA}

\author[0000-0002-0084-572X]{C. Allende Prieto}
\affiliation{Instituto de Astrofísica de Canarias, E-38205 La Laguna, Tenerife, Spain}
\affiliation{Departamento de Astrofísica, Universidad de La Laguna, E-38206 La Laguna, Tenerife, Spain}

\author[0000-0002-1691-8217]{Rachael L. Beaton}
\affiliation{Space Telescope Science Institute, 3700 San Martin Drive, Baltimore, MD 21218, USA}

\author[0000-0002-3601-133X]{D. Bizyaev}
\affiliation{Apache Point Observatory and New Mexico State University, Sunspot, NM, 88349, USA}
\affiliation{Sternberg Astronomical Institute, Moscow State University, Universitetskiy prosp. 13, Moscow, 119234, Russia}

\author[0000-0001-9205-2307]{S. Daflon}
\affiliation{Observatório Nacional/MCTIC, R. Gen. José Cristino, 77, 20921-400, Rio de Janeiro, Brazil}

\author[0000-0001-5388-0994]{S. Hasselquist}
\affiliation{Space Telescope Science Institute, 3700 San Martin Drive, Baltimore, MD 21218, USA}

\author[0000-0002-2532-2853]{Steve Howell}
\affiliation{NASA Ames Research Center, Moffett Field, CA 94035 USA}
 
\author[0000-0003-2025-3147]{Steven R. Majewski}
\affiliation{Department of Astronomy, University of Virginia, Charlottesville, VA 22904-4325, USA}

\author[0000-0002-7549-7766]{Marc Pinsonneault}
\affiliation{Department of Astronomy, The Ohio State University, Columbus, OH 43210, USA}

\begin{abstract}
Planetary radii are derived for 218 exoplanets orbiting 161 M dwarf stars. Stellar radii are based on an analysis of APOGEE high-resolution near-IR spectra for a subsample of the M-dwarfs; these results are used to define a stellar radius-M$_{\rm K_{\rm s}}$ calibration that is applied to the sample of M-dwarf planet hosts. The planetary radius distribution displays a gap over R$_{\rm p}$$\sim$1.6–2.0 R$_{\oplus}$, bordered by two peaks at R$_{\rm p}$$\sim$1.2–1.6 R$_{\oplus}$ (super-Earths) and 2.0–2.4 R$_{\oplus}$ (sub-Neptunes). The radius gap is nearly constant with exoplanetary orbital period (a power-law slope of m=$+0.01^{+0.03}_{-0.04}$), which is different (2-3$\sigma$) from m$\sim$$-$0.10 found previously for FGK dwarfs. This flat slope agrees with pebble accretion models, which include photoevaporation and inward orbital migration.
The radius gap as a function of insolation is approximately constant over the range of S$_{\rm p}$$\sim$20–250 S$_{\oplus}$. 
The R$_{\rm p}$-P$_{\rm orb}$ plane exhibits a sub-Neptune desert for P$_{\rm orb}$$<$2d, that appears at S$_{\rm p}$$>$120 S$_{\oplus}$, being
significantly smaller than S$_{\rm p}$$>$650 S$_{\oplus}$ found in the FGK planet-hosts, indicating that the appearance of the sub-Neptune desert is a function of host-star mass.
Published masses for 51 exoplanets are combined with our radii to determine densities, which exhibit a gap at $\rho_{\rm p}$$\sim$0.9$\rho_{\oplus}$, separating rocky exoplanets from sub-Neptunes. 
The density distribution within the sub-Neptune family itself reveals two peaks, at $\rho_{\rm p}$$\sim$0.4$\rho_{\oplus}$ and $\sim$0.7$\rho_{\oplus}$. 
Comparisons to planetary models find that the low-density group are gas-rich sub-Neptunes, while the group at $<$$\rho_{\rm p}$$>$$\sim$0.7$\rho_{\oplus}$ likely consists of volatile-rich water worlds.
\end{abstract}

\section{Introduction}

The first exoplanet orbiting a solar-type star was discovered in 1995 by \citet{mayor1995} and since then, almost 6000 exoplanets have been confirmed (NASA Exoplanet Archive). More than 70$\%$ of these exoplanets were detected via planetary transits, with many having M dwarfs as their stellar hosts. M-dwarf stars form a particularly interesting stellar group for studying exoplanets, as they host $\sim$3.5$\times$ more small exoplanets than FGK dwarfs \citep{mulders2015}, and are more sensitive to detection methods such as transits and radial velocities. Furthermore, because M dwarfs remain on the main-sequence significantly longer than hotter stars, they are often considered to ``live forever" in chemical evolution models and, as such, provide extended opportunities to study long-lived habitable environments for exoplanets \citep{adams2005}. 

Exoplanets detected via the transit method can have their absolute radii determined precisely, provided that the absolute radius of the host star is well known. 
The California Kepler Survey (the CKS; \citealt{petigura2017,johnson2017}) obtained high-resolution spectra for $\sim$1300 FGK stars observed by the Kepler mission, and results obtained from this data set have been very influential in the field of exoplanets. Having high-resolution spectra for exoplanet hosts has allowed for the derivation of precise stellar effective temperatures and radii. These parameters combined with precise distances from Gaia resulted in precise planetary radii revealing a dearth of planets with radii between $\sim$1.5 and $\sim$2.0 R$_{\oplus}$, depicting the small planet radius gap (Fulton gap; \citealt{fulton2017}). 
Further works analyzing samples of K2 exoplanets around FGK stars indicated that the exoplanetary radius gap was ubiquitous, as it was also present in samples of small planets formed in other environments on the Galactic disk. 

Several studies of FGK hosts have shown that exoplanets having radii falling within the gap have a dependence on the exoplanet orbital period and host star insolation  \citep{fulton2017,fulton2018,vaneylen2018,martinez2019,macdonald2019,wu2019,tacuri2023,berger2023,ho2023}, which can provide insight into the dominant mechanism responsible for sculpting the radius gap.
\cite{martinez2019} conducted a detailed spectroscopic analysis of the CKS sample (including planet completeness corrections) and found that the radius gap exhibits a slope of $-0.11\pm0.02$ with the exoplanet orbital period, and $+0.12\pm0.02$ with insolation, in the sense that the position of the radius gap decreases with increasing orbital period, and increases with increasing insolation. A similar orbital period slope in the radius gap (of $-0.09^{+0.02}_{-0.04}$) was obtained for a smaller sample of 117 Kepler stars from asteroseismology \citep{vaneylen2018}. Such results, which were obtained for host stars of FGK-types, are in agreement with models of photo-evaporation driven by the stellar XUV radiation  (e.g., \citealt{murrayclay2009,lammer2012,owen2012,owen2013,kislyakova2013,lopez2014,jin2014,chen2016,lopez2018,wu2019,jin2018,mordasini2020}) and core-powered mass loss \citep{ginzburg2018,gupta2019,gupta2020}.
These results indicate that planets below the radius gap are primarily super-Earths with rocky cores, while those above the gap are sub-Neptunes with hydrogen-helium envelopes.

Although most studies in the literature have focused on FGK stars, exoplanets around low-mass dwarfs have been the subject of previous works.
\citet{cloutier2020} examined the occurrence rates of small exoplanets detected by Kepler and K2 orbiting mid-K to mid-M dwarf stars (with masses M$\sim$0.08 -- 0.93 M$_{\odot}$) and found a positive slope of +0.058 ± 0.022 with orbital period and a negative slope of $-$0.06$\pm$0.025 with insolation. These findings suggest that gas-poor formation mechanisms might be more relevant for planets around M dwarfs as in the models by \citet{lopez2018}. 
However, results from \citet{vaneylen2021} found a negative slope with orbital period of $-0.11^{+0.05}_{-0.04}$ for M-dwarf-hosted exoplanets (the same slope obtained by \citealt{martinez2019}), aligning with the FGK trends, and also favoring photo-evaporation or core-powered mass loss models for low-mass star hosts. Another recent result comes from the work by \citet{gaidos2024} who found a negative but weak dependence of the radius gap with orbital period (slope=$-0.03^{+0.01}_{-0.03}$), suggesting a planet formation mechanism which is relatively independent of the host star properties, that can be explained by models that include orbital migration.
Finally, \citet{luque2022} found that there is a density gap (not a radius gap) as their results are inconsistent with a bimodal radius distribution for their sample; they proposed that models that include orbital migration can explain their results.

In this paper, we further probe the low-mass stellar regime of small planet hosts by doing a homogeneous study and deriving stellar parameters for a sample of M dwarf stars, with the determination of their stellar radii, and the radii of their orbiting exoplanets, using transit depth measurements from the literature. Our spectroscopic analysis is based on the high-resolution near-infrared spectra obtained by the SDSS APOGEE survey \citep{blanton2017,majewski2017_apogee}. Using the results obtained from the APOGEE spectra, we derived a calibration of stellar radius with the absolute magnitude M$_{\rm K_{\rm s}}$ to obtain stellar radii and planetary radii for a sample of 218 exoplanets (the ``Full sample") with which we can study the radius gap of small exoplanets using a homogeneous data set. 

This paper is organized as follows. In Section 2 we present the sample of planet-hosting M dwarf stars and the exoplanets that were analyzed in this study. The determination of stellar parameters for the APOGEE sample is discussed in Section 3. In Section 4, we discuss the stellar radii determination for the Full sample and, in Section 5, we derive the planetary radii for the Full sample. In Section 6 we discuss the distribution of small exoplanet radii around the sample of M dwarfs.
Finally, Section 7 summarizes the conclusions.

\section{The Sample}

\textbf{The Full Sample}: To define the sample of M dwarf hosting planets in this study, we extracted the table of confirmed planets of NASA Exoplanet Archive (meaning Kepler planets with ``Disposition'' field set as ``CP'' or ``KP'', K2 planets with ``Archive Disposition'' field set as ``CONFIRMED'', and TESS planets with ``TFOPWG Disposition'' field set as ``KP''), which by the time of download had a total of 5,806 confirmed exoplanets around 4,341 host stars. We then cut the sample to consider only stars with reported T$_{\rm eff}<$ 4100 K, $\log{g} > $4 (resulting in 445 stars), having M$_{\rm K_{\rm s}}$ absolute magnitudes 4.98 $<$ M$_{\rm K_{\rm s}}<$ 7.51 (the same magnitude range of the APOGEE sample that will be discussed below), and with distances reported in \citet{bailerjones2021}, resulting in 285 stars.
We then removed 66 stars that are known from the literature to orbit binary stars or have Gaia DR3 RUWE $>1.4$ \citep{belokurov2020}.
This initial sample has a total of 219 M dwarf stars.

We then applied another cut to consider only exoplanets with orbital periods less than 100 days, and having available transit depth measurements with transit depth errors $<$25$\%$ of the transit depth value in the NASA exoplanet archive; this resulted in a sample with 262 exoplanets.
We note, however, that the median of the fractional uncertainties in transit depths for the sample is: 5.1$\%$ with a MAD of 2.4$\%$, with most (all, except for nine) transit depth uncertainties being less than 15$\%$ of the transit depth values, which corresponds to planetary radii fractional errors of less than $\sim$8$\%$.
Finally, in this study, we consider only small planets with radii less than 4 R$_{\oplus}$ (based on the planetary radii derived in Sections 4 and 5). The large planets in the initial sample have been included in the sample analyzed in \citealt{wanderley2025}).

In summary, the Full sample analyzed here contains 161 planet-hosting M dwarf stars that have 218 small exoplanets detected by the missions Kepler (\citealt{koch2010,borucki2010,batalha2013}); 65 exoplanets), K2 (\citealt{howell2014}); 67 exoplanets), and TESS (Transiting Exoplanet Survey Satellite; \citealt{ricker2015}); 86 exoplanets). 
We note that our Full sample includes 24 M dwarfs with ongoing or scheduled spectroscopic follow-up observations with the James Webb Space Telescope (JWST). These stars are part of archival or planned JWST programs employing the NIRSpec, NIRISS, NIRCam, and MIRI instruments to characterize transiting exoplanets.
The results for the Full M dwarf sample are in Table \ref{fulldata}, and will be discussed in Section 6.

\textbf{The Apogee Sample:}
This work is based on the analysis of near-infrared ($\lambda$1.51 $\mu$m to $\lambda$1.69 $\mu$m) high-resolution (R$\sim$22,500) spectra of M dwarf stars from the APOGEE survey \citep{majewski2017_apogee}. The APOGEE spectra analyzed were obtained using two 2.5-m telescopes, one located at Apache Point Observatory (APO) in the northern hemisphere and the other at Las Campanas Observatory (LCO) located in the southern hemisphere \citep{bowen1973,gunn2006_sdss,wilson2019}.

To select M dwarfs planet hosts that had been observed by APOGEE, we cross-matched the sample of 161 M dwarf stars from above with the SDSS APOGEE DR17 \citep{apogeedr17_2022}, requiring that the APOGEE spectra have S/N $>$ 50, finding 42 stars. To increase the sample, we added 3 M dwarf stars that have exoplanets detected by radial velocities, and 3 M dwarfs that host planets with radii larger than 4 R$_{\oplus}$, which are studied in \citet{wanderley2025}.
The APOGEE sample analyzed contains 48 M dwarf stars that have confirmed exoplanets, and their results are presented in Table \ref{fulldata}.

\section{Stellar Parameter Determination for the APOGEE Sample}

The derivation of fundamental atmospheric parameters for the sample M dwarfs was performed via spectrum synthesis analyses, using the APOGEE DR17 line list \citep{smith2021}, along with MARCS stellar model atmospheres \citep{gustafsson2008_marcs}, and the radiative transfer code Turbospectrum \citep{plez2012_turbospectrum}. 

As discussed in \citet{souto2018}, the OH lines in the H-band spectra of M dwarfs exhibit a weak sensitivity to the effective temperature, while H$_{2}$O lines are highly sensitive to T$_{\rm eff}$, becoming much weaker for hotter M dwarfs compared to cooler M dwarfs.  
The stellar effective temperatures (T$_{\rm eff}$), surface gravities ($\log{g}$), metallicities ([M/H]), as well as the oxygen abundances and projected rotational velocities (v$\sin{i}$), were derived from best fits between the APOGEE spectra and model spectrum syntheses, using different wavelength windows or spectral features for the derivation of each set of parameters (see \citealt{wanderley2023,wanderley2024b} for details).

We selected spectral regions containing mainly H$_{\rm 2}$O and OH lines and searched for the T$_{\rm eff}$ -- A(O) and $\log{g}$ -- A(O) pairs that best fit the observed spectra (see \citealt{wanderley2023} for details). The following spectral windows were selected: $\lambda$15253.3 -- 15274.1 \r{A}; $\lambda$15312.2 -- 15321.0 \r{A}, $\lambda$15350.1 -- 15364.0 \r{A}; $\lambda$15369.1 -- 15376.1 \r{A}; $\lambda$15403.7 -- 15412.7 \r{A}; $\lambda$15444.1 -- 15465.0 \r{A}; $\lambda$15500.1 -- 15509.3 \r{A}; $\lambda$15554.6 -- 15575.6 \r{A}. 

\begin{figure*}
\begin{center}
  \includegraphics[angle=0,width=0.8\linewidth,clip]{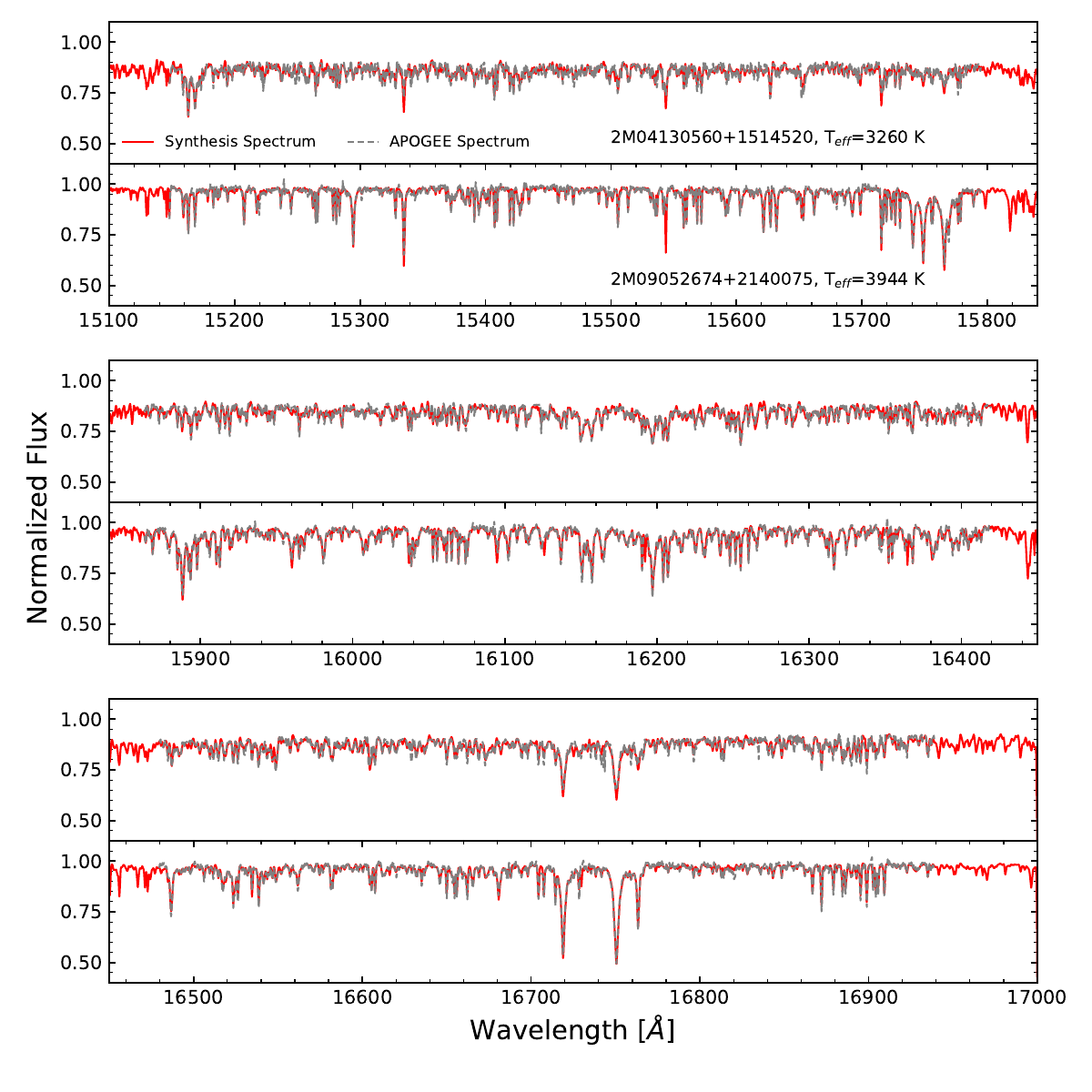}
\caption{The observed APOGEE spectrum and best-fit syntheses for two studied M dwarfs, covering the effective temperature range of our sample. From top to bottom: the first, second, and third chip of the APOGEE spectrum are shown. 
}
\end{center}
\label{sint_obs}
\end{figure*}

Figure \ref{sint_obs} shows, as an example, the best-fit syntheses (red lines) for the observed APOGEE spectra (grey dashed lines) for two target M dwarf stars. The top and bottom portions of each panel of the figure show, respectively, the spectrum of and 2M04130560+1514520 (a fully convective M dwarf; T$_{\rm eff}$ = 3260 K); and 2M09052674+2140075 (a partially convective M dwarf; T$_{\rm eff}$ = 3944 K) these two targets span the range of effective temperatures in this study. 

\begin{figure}
\begin{center}
  \includegraphics[angle=0,width=1\linewidth,clip]{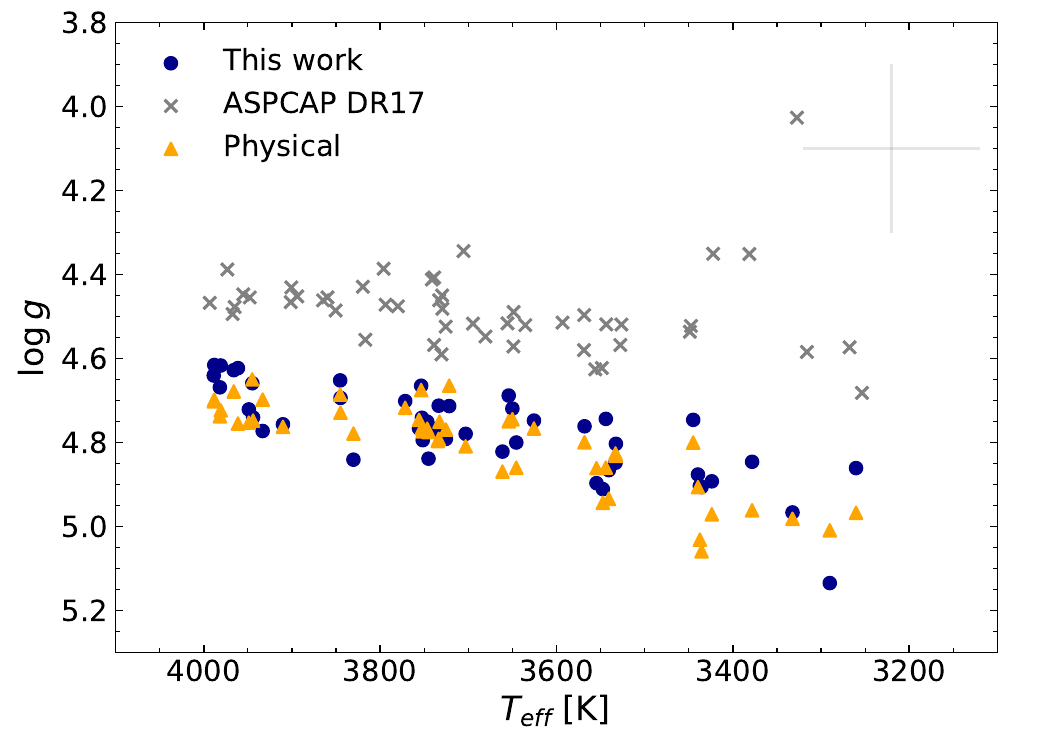}
\caption{Kiel Diagram showing the effective temperatures and surface gravities derived in this study (blue circles), along with ASPCAP DR17 results for the same stars (grey xs). Physical $\log{g}$ values computed using the relations in \citet{mann2019} are shown as orange triangles. Typical uncertainties are shown at the top right. The DR17 $\log{g}$ results are systematically lower than ours and roughly flat with the effective temperature.}
\end{center}
\label{kiel}
\end{figure}

The stellar parameters derived in this study for the APOGEE sample are shown in the Kiel diagram of Figure \ref{kiel} as blue circles. These results confirm that the sample targets are M dwarf stars. Their effective temperatures and surface gravities are in the range between $\sim$3200 -- $\sim$4000 K, and $\sim$4.6 -- 5.1, respectively. 
Figure \ref{kiel} also shows a comparison of the surface gravities in this work with physical $\log{g}$ values (orange triangles), the latter were determined from the derived stellar radii, along with stellar masses obtained using photometric calibrations from \citet{mann2019}. Overall, there is good agreement between both sets of $\log{g}$ values, with an average offset (and STD) of $-0.03$ $\pm$ 0.06.

The uncertainties in the derived parameters were estimated in detail in our previous studies and these are: $\pm$100 K for T$_{\rm eff}$ and $\pm$0.2 dex for $\log{g}$ \citep{souto2018,souto2020}. 
In summary, these uncertainties were estimated by assuming that $\delta$(A(OH) – A(H2O)) can differ by the typical measurement precision of $\pm$0.10 dex, resulting in the quoted uncertainties in T$_{\rm eff}$ and $\log{g}$ (see the right panels of figure 1 from \citealt{souto2020}).
As in the previous works, our methodology to derive stellar parameters uses the sensitivities of OH and H$_{\rm 2}$O lines to oxygen abundance determinations as a function of T$_{\rm eff}$ and $\log{g}$ to derive T$_{\rm eff}$-A(O) and $\log{g}$-A(O) pairs simultaneously.
\citet{souto2018} also investigated the stellar parameter sensitivity to changing the model atmospheres from MARCS to PHOENIX BT-Settl \citep{allard2013} models, finding that the differences in the obtained parameters are very small (see Table 1 from \citealt{souto2018}). 

\begin{figure}
\begin{center}
  \includegraphics[angle=0,width=1\linewidth,clip]{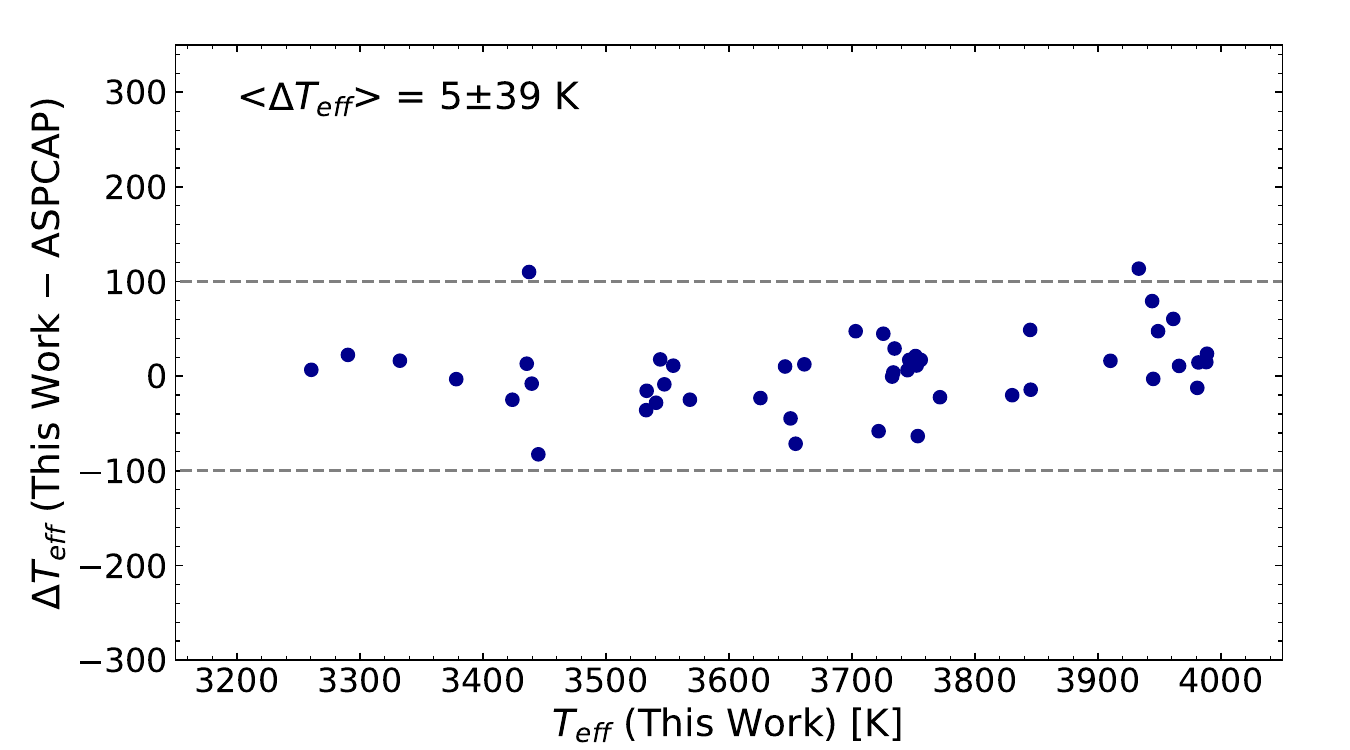}
\caption{A comparison between the derived effective temperatures from this work with results obtained with the ASPCAP pipeline from APOGEE DR17. The dashed lines represent offsets of $\pm$ 100 K.}
\end{center}
\label{teffcomp}
\end{figure}

For comparison, we also show in Figure \ref{kiel} the ASPCAP DR17 (Turbospectrum 2020 release) results for the same stars (grey xs), which were derived with the ASPCAP pipeline \citep{garciaperez2016_aspcap}. It is clear that the log g values from ASPCAP DR17 are systematically smaller than ours, having a roughly constant value with the effective temperature, while ours tend to decrease with T$_{\rm eff}$, as expected. 
In Figure \ref{teffcomp} we show the effective temperatures from this work versus those from DR17. Our results are found to be in excellent agreement with those of ASPCAP, having only a small average offset of $6\pm38$ K. The differences between the effective temperatures (This Work - ASPCAP) for all stars, except two, are within $\pm$100 K (dashed line in the figure), but there is a small trend for effective temperatures larger than $\sim$3800 K. 

Stellar radii for the APOGEE sample were obtained from the Stefan-Boltzmann equation, and using the effective temperatures from Section 3. The derivation of stellar luminosities relied on Gaia stellar distances from \citet{bailerjones2021}, along with calibrations for 2MASS J-band bolometric corrections from \citet{mann2015,mann2016}, with J and V magnitudes collected from \citet{gsc2008,zacharias2012ucac4,furlan2017}, and \citet{skrustkie2006}. We used dust maps from \citet{green2018} to de-extinct the stellar magnitudes, although this is a nearby sample of M dwarfs. Bolometric magnitudes were converted into stellar luminosities adopting the zero-point luminosity L$_{0}$ of 3.0128 $\times$ 10$^{35}$ erg s$^{-1}$ from \citet{mamajek2015}, along with the equation below:

\begin{equation}
    L_{*}=L_{0} 10^{-0.4M_{\rm bol}}
    \label{aaa}    
\end{equation}

The derived luminosities span the range between 0.006 and 0.085 L$_{\odot}$, and the radii span the range between $\sim$0.2 -- $\sim$0.6 R$_{\odot}$.  
The stellar radii errors were derived from the propagation of the luminosity uncertainties (derived from the distance and bolometric correction uncertainties from respectively \citealt{bailerjones2021} and \citealt{mann2015,mann2016}) and considering effective temperature uncertainties of $\pm$100 K. The ensemble of stellar parameters obtained for the APOGEE stars (effective temperatures, surface gravities, M$_{\rm K_{\rm s}}$ absolute magnitudes, luminosities and stellar radii) are presented in Table \ref{fulldata}.
The parameters [M/H], A(O), and projected rotational velocities (v$\sin{i}$) are published in a companion paper \citet{wanderley2025} that studies the metallicities and oxygen abundances in planet-hosting stars.

\section{Stellar Radii for the Full Sample}

\begin{figure}
\begin{center}
  \includegraphics[angle=0,width=1\linewidth,clip]{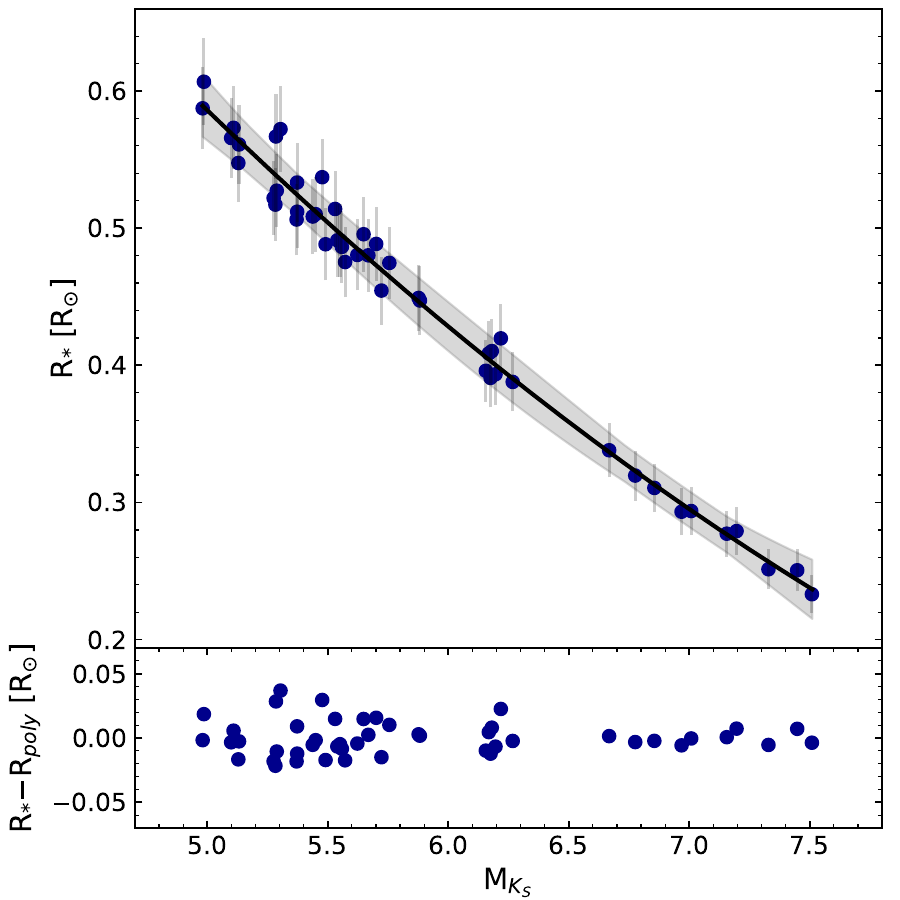}
\caption{Top panel: stellar radii as a function of M$_{\rm K_{\rm s}}$ for our APOGEE sample are shown as blue circles. The second-degree polynomial fit for our data is represented as a black line, and is given by R$_{*}$=1.7420 $-$0.2925 M$_{\rm K_{\rm s}}$ + 0.0123 M$_{\rm K_{\rm s}}^{2}$. The gray region shows the derived uncertainties in stellar radii as a function of M$_{\rm K_{\rm s}}$. The bottom panel shows the residuals between our derived radii and the calibration.}
\end{center}
\label{calib}
\end{figure}

Using the spectroscopic parameters and stellar radii obtained previously in this study, we derived a calibration between the M dwarf stellar radius and the M$_{\rm K_{\rm s}}$ absolute magnitude. In Figure \ref{calib} we show the results for the APOGEE sample (filled blue circles) and the best-fit relation obtained (solid black line). We found that a second-degree polynomial represented the data well such as:

\begin{equation}
    R_{*}=a_{0} + a_{1} M_{\rm K_{\rm s}} + a_{2} M_{\rm K_{\rm s}}^{2}
    \label{rmk}    
\end{equation}

\noindent where the coefficients are: a$_{0}$=1.7420, a$_{1}$=$-$0.2925, and a$_{2}$=0.0123, and their standard deviations are, respectively, 0.1599, 0.0522, and 0.0042. 
We applied the calibration above to obtain stellar radii homogeneously for the Full sample of M dwarfs.

Below is the covariance matrix of the polynomial coefficient estimates in 10$^{-4}$ units:

\[
\begin{pmatrix}
255.7663 & -83.4253 & 6.6959 \\
-83.4253 & 27.2795 & -2.1951 \\
6.6959 & -2.1951 & 0.1771
\end{pmatrix}
\]

The uncertainties in the stellar radii obtained using the calibration are depicted as the grey region in the top panel of Figure \ref{calib}. These were derived by considering errors in stellar radii as a function of M$_{\rm K_{\rm s}}$, which is related to the polynomial coefficient estimates, and also considering the residuals between the fit and the data. The uncertainties related to the polynomial coefficient estimates were obtained by drawing random samples from a multivariate normal distribution, and combining the stellar radii errors of all of these. 
For each realization, there is a different set of coefficients and a covariance matrix, which are used to derive stellar radii errors as a function of M$_{\rm K_{\rm s}}$.
Figure \ref{calib} illustrates the scatter between the input data and the fitted polynomial, with the bottom panel showing the residuals of $\Delta$R=R$_{\rm *}-$R$_{poly}$. A quadratic polynomial was fit to the absolute values of (R$_{\rm *}-$R$_{poly}$) as a function of M$_{\rm K_{\rm s}}$ and were added in quadrature with the uncertainties related to the polynomial coefficient estimates to represent total uncertainties in R$_{\rm *}$. The shaded region in the top panel of Figure \ref{calib} defines these uncertainties as a function of M$_{\rm K_{\rm s}}$.

The previous work by \citet{souto2020} also derived a calibration for M$_{\rm K_{\rm s}}^{2}$ and stellar radius based on their results from a spectroscopic analysis of APOGEE spectra of 21 benchmark M dwarfs. To compare the two calibrations, which are based on slightly different spectroscopic methodologies and line lists, we used their calibration to derive stellar radii for our sample and found a small percentage difference between the results of $\sim$2$\%\pm$0.7$\%$, indicating good agreement between the relations.

\begin{figure}
\begin{center}
  \includegraphics[angle=0,width=1\linewidth,clip]{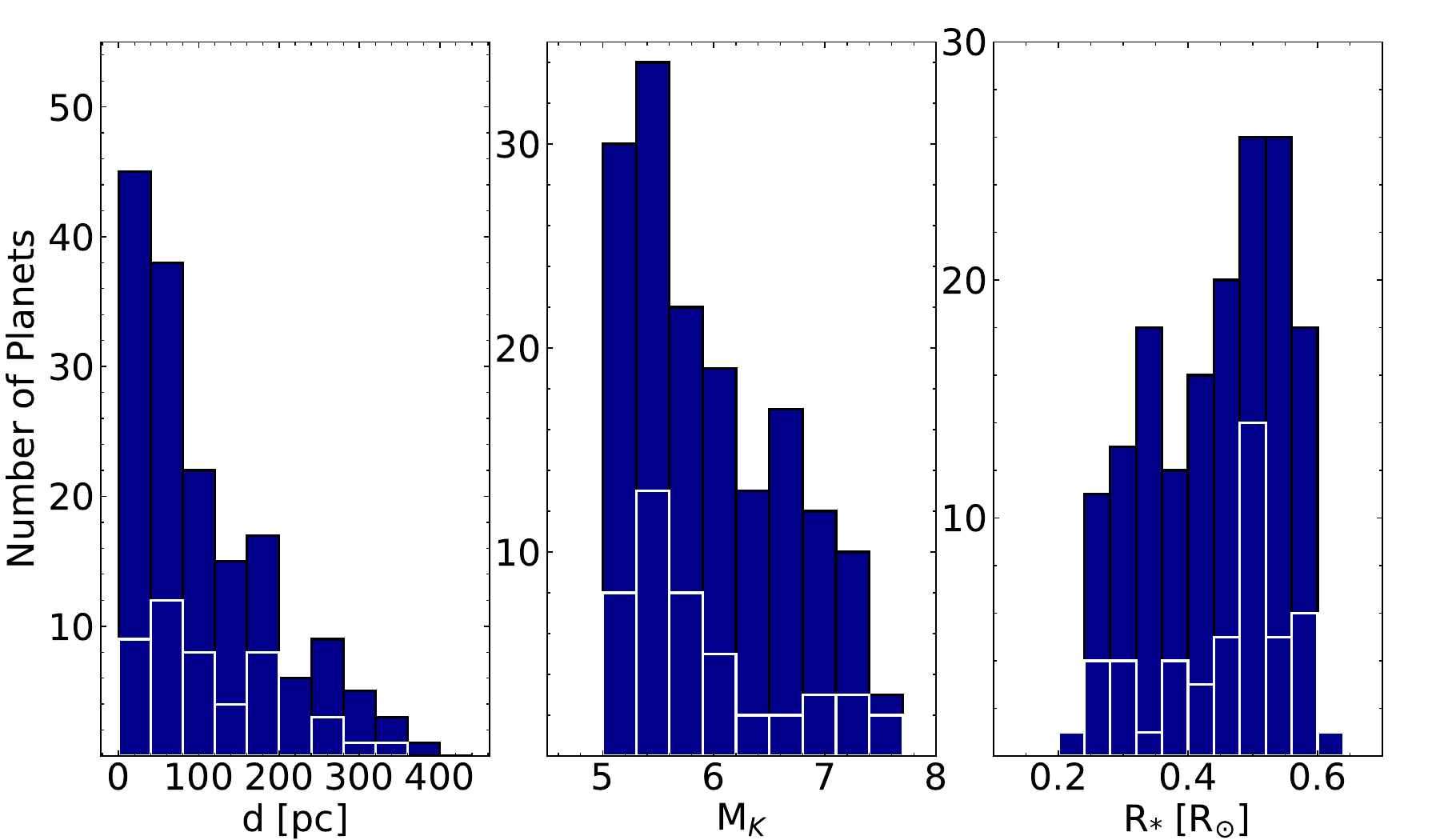}
\caption{From left to right: distribution of Gaia distances from \citet{bailerjones2021}, 
M$_{\rm K_{\rm s}}$ absolute magnitudes, and 
stellar radii for the 161 M dwarfs in our Full sample. As a comparison we also show the histograms for the APOGEE sample delineated in white.}
\end{center}
\label{distance}
\end{figure}

Figure \ref{distance} (right panel) shows the distribution of stellar radii for the Full sample, while the middle panel shows the absolute magnitudes $M_{\rm K_{\rm s}}$, and the left panel shows the Gaia EDR3 \citet{bailerjones2021} distances. As expected, given that M dwarfs are not luminous, the sample stars are rather nearby, having distances less than $\sim$ 380 pc, and thus represent a stellar population within the solar vicinity.
For comparison, we also show in Figure \ref{distance} corresponding histograms for the APOGEE sample delineated in white.
We can see that The APOGEE and the Full sample cover roughly the same range in parameters.

\section{Exoplanetary Radii Determination}

The stellar radii for the Full sample (obtained homogeneously from Equation \ref{rmk}) were used in conjunction with planetary transit depths to compute the radii for 218 exoplanets. 
The transit depth ($\Delta$F) is given as the ratio between the cross-section of the planet and the star, and the exoplanet radius can be derived using the equation below:

\begin{equation}
    R_{p}= \Delta F^{0.5} \times R_{*}
    \label{eqtransit}    
\end{equation}

Transit depth values for the studied sample were taken from the Kepler Objects of Interest Catalog (KOI, \citealt{burke2014,rowe2015,mullally2015, coughlin2016,thompson2018}), the ExoFOP TOI Program (Exoplanet Follow-up Observing Program - TESS Objects of Interest), as well as other works from the literature with transit depth measurements reported in the NASA Exoplanet Archive.
When more than one transit depth measurement was available, the median $\Delta$F was adopted. 
Planetary radii errors were estimated from the propagation of the errors in transit depth and stellar radius. 
We obtained the respective orbital periods and semi-major axis and their uncertainties from the NASA Exoplanet Archive. 

\begin{figure}
\begin{center}
  \includegraphics[angle=0,width=0.9\linewidth,clip]{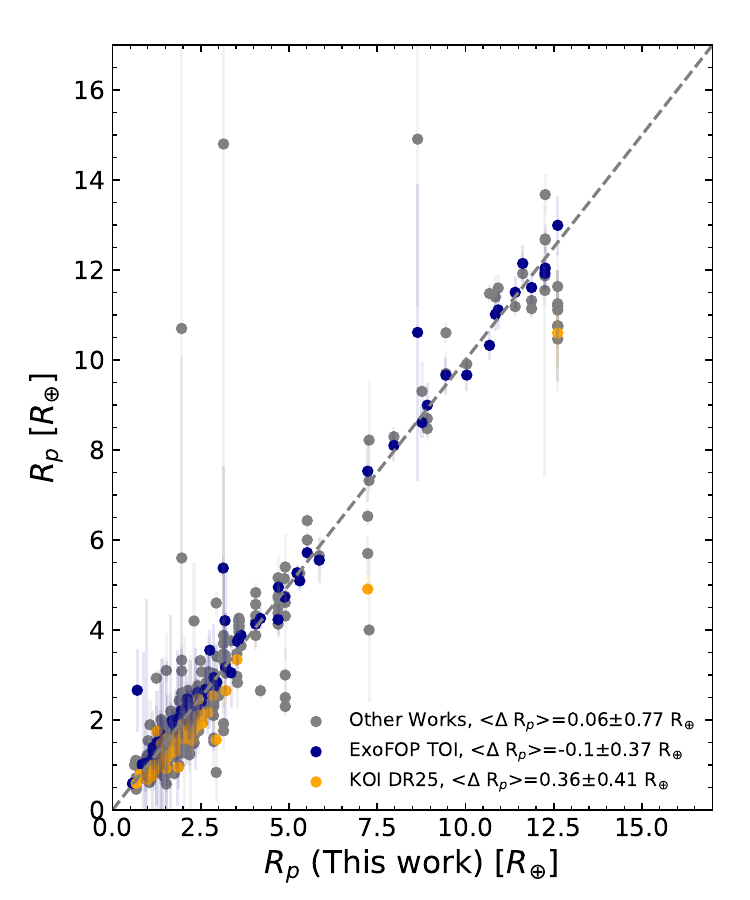}
\caption{Comparison between the planetary radii from this work with results from the literature. Blue, orange, and grey circles are radii obtained from the ExoFOP TOI Program, KOI DR25 \citep{thompson2018}, and other works from the literature. The x=y relation is shown by the grey dashed line.} 
\end{center}
\label{litcomp}
\end{figure}

In Figure \ref{litcomp} we show a comparison of the planetary radii obtained in this work with planetary radii taken from the literature. Comparisons with radii from the ExoFOP TOI Program and KOI DR25 \citep{thompson2018} are shown as the filled blue and orange circles, respectively, while the filled grey circles are comparisons with other works of the literature reported in the NASA Exoplanet Archive Database. Our radii are systematically larger than those in KOI DR25, with average (mean $\pm$ STD) differences $<$$\Delta$ R$_{\rm p}$$>_{\rm This \, Work - KOI \, DR25}$ = 0.36 $\pm$ 0.41 R$_{\oplus}$ (standard error=$\pm$0.05 R$_{\oplus}$). The planets with radii from ExoFOP TOI have radii smaller than ours: $<$$\Delta$ R$_{\rm p}$$>_{\rm This \, Work - EXOFOP}$ = $-$0.1 $\pm$ 0.37 R$_{\oplus}$ (standard error=$\pm$0.03 R$_{\oplus}$). Finally, the average difference with other works is $<$$\Delta$ R$_{\rm p}$$>_{\rm This \, Work - Other \, Works}$ = 0.06 $\pm$ 0.77 R$_{\oplus}$ (standard error=$\pm$0.03 R$_{\oplus}$). For all cases, the standard deviation is significant.

\section{Discussion}

\subsection{The Distribution of Small Exoplanet Radii Around the M dwarf Sample}

\begin{figure}[H]
\begin{center}  \includegraphics[angle=0,width=1\linewidth,clip]{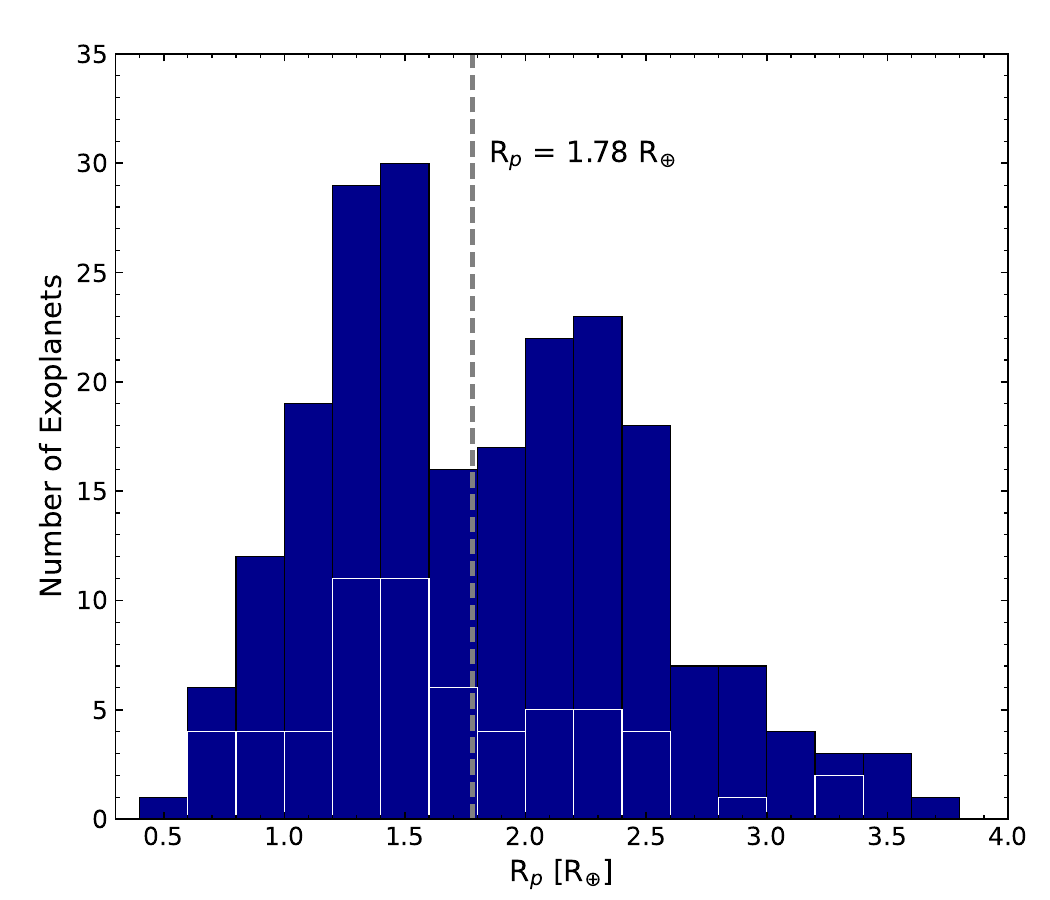}
\caption{Distribution of the derived planetary radii for our Full sample is shown in blue. The grey dashed line indicates the minimum probability density at 1.78 R$_{\oplus}$ for a KDE associated with the Full sample.
The histogram of the planetary radii distribution, obtained from the spectroscopic analysis of the APOGEE sample, is delineated in white.}
\end{center}
\label{rplhist}
\end{figure}

Figure \ref{rplhist} shows a histogram with the distribution of the derived exoplanet radii for the Full sample in blue. 
The grey dashed line at R$_{\rm p}$=1.78 R$_{\oplus}$ indicates the location with the lowest probability density for a KDE (kernel density estimation) applied to the exoplanet radius distribution of the Full sample. 
There is a clear rocky peak in the distribution of exoplanet radii between $\sim$1.2 -- 1.6  R$_{\oplus}$ in the region of the Super-Earths, with a drop in exoplanet counts for R$_{\rm p}$$<$1.2, which may be the result of biases due to more inefficient detection of very small planets. For larger exoplanets, the R$_{\rm p}$ distribution also shows a clear drop in exoplanet counts for R$_{\rm p}$ between 1.6 and 2.0 R$_{\oplus}$, and a subsequent raise in the distribution for radii between 2.0 and 2.6 R$_{\oplus}$ in the region of the sub-Neptunes (non-rocky peak). 
Although we do not apply completeness corrections to our results, in part because we are studying exoplanets detected by three different missions (Kepler, K-2 and TESS), we note that, for our sample, the sub-Neptune peak is smaller but not significantly smaller than the rocky peak: 112 exoplanets from our sample are on the rocky side as opposed to 106 being on the non-rocky side. 

For completion, we also show in Figure \ref{rplhist} the distribution of exoplanetary radii obtained from spectroscopy for the APOGEE sample (histogram delineated in white). For this smaller sample, we also find that the distribution of radii has a peak at $\sim$1.2 – 1.6 R$_{\oplus}$, a minimum at roughly the same location as the one found for the Full sample, but there is only a very small accumulation of planets in the sub-Neptune region at 2.0 – 2.6 R$_{\oplus}$ for this sample.

\citet{cloutier2020} computed occurrence rates for Kepler and K2 planets and found that the relative occurrence rate of rocky to ``non-rocky" planets greatly increases from K to M hosts. For the stellar mass range between 0.08 $>$ M $>$ 0.65 M$_{\odot}$ they find a rocky peak and a much smaller non-rocky peak at 2 R$_{\oplus}$, with a hint of a gap between 1.6 -- 1.8 R$_{\oplus}$. Overall, our results are similar, but our sample contains a proportionally larger population of planets with radii between $\sim$2 and 2.5 R$_{\oplus}$, which may be explained by the absence of completeness corrections in this study. 
Concerning other radius gap results for M dwarf hosts from the literature, \citet{luque2022} do not find evidence for a gap in their exoplanet sample (See further discussion on exoplanet densities in Section 6.3.), while the recent study by \citet{gaidos2024} finds an exoplanet radius distribution similar to ours. (see also \citet{hirano2018}).
Finally, the radius gap for M dwarf hosts obtained here ($\sim$1.6 -- 2.0 R$_{\oplus}$) is in line with the radius gap for the more massive FGK hosts \citep{fulton2017, vaneylen2018, martinez2019,petigura2022,tacuri2023}. 

\subsection{The Slope in the Radius Gap: Orbital Period and Insolation}

\subsubsection{Orbital period}
 
In this study, we use a combined sample of exoplanetary radii and orbital periods from three different missions, Kepler, K2 and TESS. 
Although the three missions have different sampling time intervals, a straightforward examination of our sample reveals similar distributions of exoplanetary orbital periods, except for eight very short orbital period exoplanets, with P$_{\rm orb}$$<$0.7 days, that were found by TESS.
For longer orbital periods, our sample has 66 exoplanets with P$_{\rm orb}\geq$10 days, with 32$\%$, 30$\%$ and 38$\%$ of these being, respectively, from Kepler, K2 and TESS. There are 14 exoplanets with  P$_{\rm orb}\geq$25 days, with four from Kepler, four from K2 and six from TESS.

\begin{figure}
\begin{center}
  \includegraphics[angle=0,width=0.85\linewidth,clip]{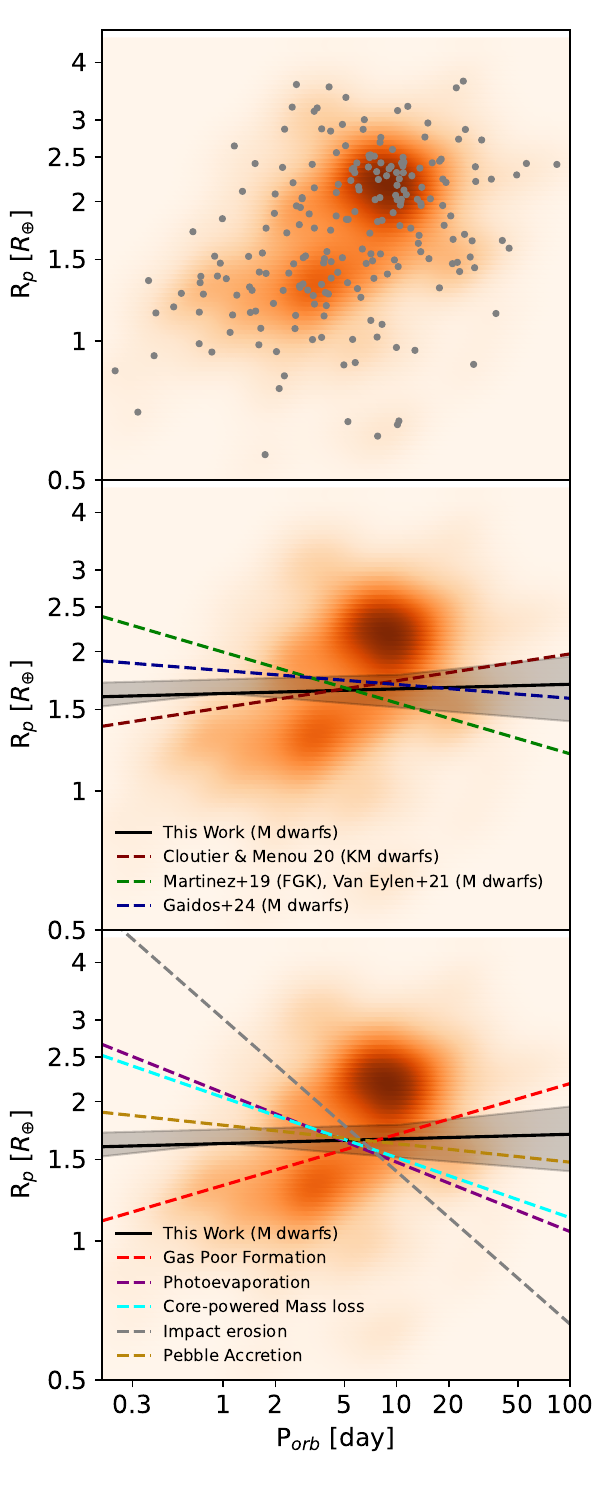}
    \caption{The top panel shows the distribution of the derived planetary radii as a function of orbital periods along with the associated KDE. The middle and bottom panels show the same KDE, along with our radius gap relation (black solid line), and uncertainties (grey region). Middle panel: the radius gap relation from \citet{cloutier2020} for KM dwarfs is depicted as a maroon dashed line, the relation from \citet{martinez2019} for FGK dwarfs and \citet{vaneylen2018} for M dwarfs are depicted as a green dashed line, and the relation from \citet{gaidos2024} for M dwarfs is depicted as a blue dashed line. Bottom panel: prescriptions from different models. Red, purple, cyan, grey, and orange dashed lines represent, respectively, models for gas-poor formation \citep{lopez2018}, photo-evaporation \citep{lopez2018}, core-powered mass loss \citep{gupta2019}, impact erosion \citep{wyatt2020}, and pebble accretion \citep{venturini2024}.}
\end{center}
\label{porbr}
\end{figure}

As discussed previously, the gap in the distribution of exoplanetary radii for FGK stellar hosts is dependent on the orbital period P$_{\rm orb}$ (in days) and on the insolation S$_{\rm p}$ (in Earth fluxes S$_{\oplus}$; \citealt{martinez2019,vaneylen2018}), with relations defined by slopes between these variables on a logarithmic scale: dlog(R$_{\rm p}$)/dlog(P$_{\rm orb}$) and dlog(R$_{\rm p}$)/dlog(S$_{\rm p})$. To investigate the slope in the exoplanet radius gap for the exoplanets orbiting our M dwarf sample, we applied a KDE and extracted probability densities for the parameter space of our sample. Given these probabilities, we simulated lines for different slopes and intercept pairs and summed the densities of the points across the trajectory of each model. The line with the lowest probability density summation was taken as the radius gap slope and intercept. We then used this slope as input to the \textit{gapfit} code \citep{loyd2020,berger2023}, bootstrapping our sample 1000 times, and defining the line that best represents the gap in our M dwarf data. Finally, we adopted the average slope between the first fit (slope=$-0.01$) and the subsequent \textit{gapfit} one (slope=$+0.02$). This resulted in a slope of m = $+0.01^{+0.03}_{-0.04}$ and an intercept of y0 = $+0.21^{+0.03}_{-0.01}$, or R$_{\rm gap}$=1.62R$_{\rm p}$P$_{\rm orb}^{+0.01}$. 
The upper panel of Figure \ref{porbr} shows the derived planetary radii as a function of the exoplanet orbital periods (grey circles) along with a KDE of the planet density in this parameter space. A visual inspection of this figure shows the presence of the gap, with a hint of possibly quite flat slope with orbital period. 
In the middle and bottom panels of this figure we show our KDE, along with the radius gap relation with the orbital period obtained in this work (black solid line). 
The uncertainty band in this relation is depicted as a grey region. For comparison, we also show in the middle panel of Figure  \ref{porbr} results from other works: \citealt{cloutier2020} (maroon dashed line), \citealt{martinez2019} (green dashed line), \citealt{vaneylen2021} (superposed with the slope from \citealt{martinez2019} for FGK hosts), and \citealt{gaidos2024} (blue dashed line).

Figure \ref{porbr} (bottom panel) shows the slopes for planet formation models from the literature: 1) gas-poor formation model (\citealt{lopez2018}, red dashed line), which implies late gas accretion for a rocky exoplanet population when the gaseous protoplanetary disk has almost dissipated; 2) photoevaporation model (\citealt{lopez2018}, purple dashed line), which implies an atmospheric loss for close-in planets due to photoevaporation caused by EUV and X-ray radiation from a young star; 3) core-powered mass loss model (\citealt{gupta2019}, cyan dashed line), which implies atmospheric escape due to a cooling core radiating its formation energy; 4) impact erosion model (\citealt{wyatt2020}, grey dashed line), which implies either atmospheric stripping or the formation of a secondary volatile atmosphere due to planetesimal impacts that depends on the planet's size and distance from the star; 5) pebble accretion model (\citealt{venturini2024}, orange dashed line), which includes the effects of photoevaporation along with pebble accretion, and predicts a near-flat slope.

Comparisons between observational results and predictions from models of planetary formation (middle and bottom panels of Figure \ref{porbr}) can offer insight into the mechanism responsible for sculpting the radius gap. For FGK stars, the slope from \citet{martinez2019} aligns well with predictions from photoevaporation (\citealt{lopez2018}) and core-powered mass loss (\citealt{gupta2019}) models. 
We point out, however, that several other works, e.g., \citet{vaneylen2018}, \citet{ho2023}, and \citet{petigura2022}, obtained similar slopes ($\sim$$-$0.10) for this stellar class, making it a well established result that photoevaporation /core-powered mass loss are the dominant mechanism in the formation of the small planet radius gap in solar-type stars.

For M dwarf hosts, the results from the literature vary between a negative 
slope=$-0.11^{+0.05}_{-0.04}$ \citep{vaneylen2021}, aligning well with photoevaporation / core powered mass loss models, and a positive 
slope=$+0.058^{+0.022}_{-0.022}$ \citep{cloutier2020}, which favors gas-poor formation models. The range in stellar mass in these two studies is similar: 0.08 -- 0.93 M$\odot$ for \citet{cloutier2020} and 0.09 -- 0.65 M$\odot$ for \citep{vaneylen2021}, and it seems unlikely that different mass ranges can easily explain the very different slopes. 
The study by \citet{gaidos2024} found a quasi-flat slope within the uncertainties (slope=$-0.03^{+0.01}_{-0.03}$) for their sample of M dwarfs, while
a negative slope ($-0.065^{+0.024}_{-0.013}$) was obtained by the recent work by \citealt{bonfanti2024}, 
who interpreted their less steep slope (when compared with $\sim$-0.11) as resulting from a sample composed of a mixture of exoplanets that suffered photo-evaporation / core-powered mass loss and gas-poor formation. 
As recognized in \cite{vaneylen2021}, it is possible that the use of heterogeneous results from the literature and small samples are the culprit for the discrepant results found in the literature for exoplanets orbiting M dwarfs. 

Our study has the advantage of having an exoplanet sample of reasonable size (compared to the current number of exoplanets that have been detected to orbit M dwarfs), and of using precise exoplanet radii that ultimately come from a detailed spectroscopic analysis of APOGEE spectra and derived homogeneously from a calibration of APOGEE results with M$_{\rm K_{\rm s}}$. 

The results for our sample of exoplanets around M dwarf hosts indicate a nearly flat slope. 
Flat slopes can be explained from a combination of photoevaporation and inward migration of icy planets \citep{luque2022,bonfanti2024}. This would transform an ``original'' negative slope produced by photoevaporation / core-powered mass loss into a flat slope. Since inward planet migration is more efficient for small planets that orbit lower-mass stars, this could explain why we obtain a more negative slope when considering exoplanets around the more massive FGK dwarf stars.

As shown in the bottom panel of Figure \ref{porbr}, most of the models predict R$_{\rm P}$ - P$_{\rm orb}$ power-law slopes that are much steeper (both negative and positive) than our derived relation for the M-dwarf exoplanets. The pebble-accretion models from \citet{venturini2024}, which consider the effects of orbital migration, as well as photoevaporation to account for atmospheric loss over time, on the other hand, predict a flatter slope of R$_{\rm gap}$ with period, in much closer agreement with our slope of $+0.01^{+0.03}_{-0.04}$. This suggests that pebble accretion may play a more important role in disks around M dwarfs than in disks around FGK stars. 

\subsubsection{Insolation}

Using the stellar luminosities (L$_{*}$) derived in the previous section, along with the exoplanet semi-major axis $a$, we computed the insolation (S$_{\rm p}$), which is the flux that the exoplanet receives from the host star, and is given by the equation below, where S$_{\oplus}$ and a$_{\oplus}$ are respectively the insolation that the Earth receives from the Sun, and the Earth's semi major axis of one astronomical unit:

\begin{equation}
    \frac{S_{\rm p}}{S_{\oplus}}=\frac{L_{*}a_{\odot}^{2}}{L_{\odot} \, a^{2}} 
    \label{sp}
\end{equation}

In Table \ref{fulldata} we list stellar radii, transit depths, planetary radii, orbital periods, insolation levels, planetary masses and planetary densities for the Full sample.

\begin{figure}
\begin{center}
  \includegraphics[angle=0,width=1\linewidth,clip]{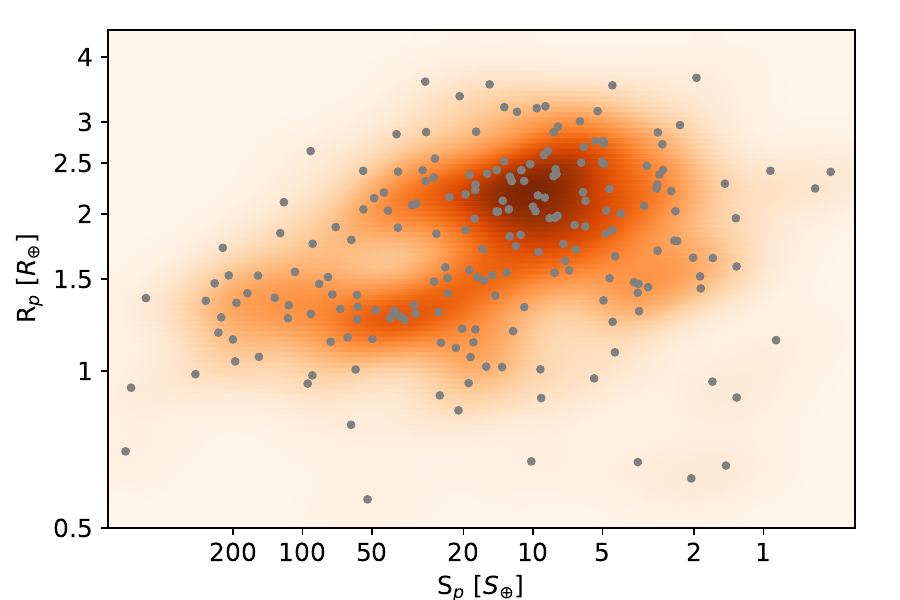}
\caption{Distribution of the derived planetary radii as a function of insolation, along with the associated KDE.} 
\end{center}
\label{insolation}
\end{figure}

Figure \ref{insolation} investigates the behavior of exoplanetary radii with incident stellar flux by plotting R$_{\rm p}$ as a function of insolation, S$_{\rm p}$, which is given in units of the insolation for the Earth, S$_{\oplus}$. This figure also shows the associated KDE.  
One of the apparent coherent features in Figure \ref{insolation} is the gap in the distribution of exoplanets having R$_{\rm p}\sim$1.6--2.0 R$_{\oplus}$ in the insolation interval of roughly S$_{\rm p}>$ 20 S$_{\oplus}$. This radius interval corresponds to the radius gap that is observed in the frequency histogram of radii and here this radius gap is mapped into insolation space.

One important feature that can be seen in both Figures \ref{porbr} and \ref{insolation} is the presence of a ``sub-Neptune desert'' (e.g., the lack of sub-Neptunes at short orbital periods and high insolation levels due to photoevaporation from the host star). The figures show that our sub-Neptunes (2 R$_{\oplus}<$R$_{\rm p}<$4 R$_{\oplus}$) are shifted overall toward lower insolation levels and longer orbital periods, with an abscence of sub-Neptunes with P$_{\rm orb}\lesssim$2 days, if compared to the rocky planets ($R_{\rm p}<2 R_{\oplus}$).

A critical point concerning the sub-Neptune desert in Figure \ref{insolation} is that its edge is found at S$_{\rm p}\sim$120 S$_{\oplus}$, which is a significantly lower value of insolation than this edge in the more massive FGK exoplanet-hosting stars, as found by \citet{lundkvist2016} to fall at S$_{\rm p}\sim$650 S$_{\oplus}$. \citet{lundkvist2016} analyzed an asteroseismically-selected sample of 102 Kleper planet-hosting stars, with most host masses falling between 0.9 to 1.6 M$_{\odot}$, which is significantly larger than the mass range of M dwarfs analyzed here, with M$\sim$0.2--0.6 M$_{\odot}$. The insolation difference between 650 S$_{\oplus}$ for a sample of 0.9 to 1.6M$_{\odot}$ planet-hosting stars compared to $\sim$120 S$_{\oplus}$ for 0.2 to 0.6 M$_{\odot}$ planet hosts suggests that the sub-Neptune desert is shaped, to some degree, by the properties of the host star. The lower values of insolation for the edge of the sub-Neptune for low-mass planet-hosting stars have also been noted by \citet{hirano2018} and \citet{gaidos2024}.

\subsection{Trends with Exoplanet Density}

The next step beyond examining the exoplanetary radii themselves is to combine our derived radii with values of exoplanetary masses taken from literature studies in order to compute mean densities, in units of the mean density of the Earth, ($\rho_{\rm p}$/$\rho_{\oplus}$)= (M$_{\rm p}$/M$_{\oplus}$)/(R$_{\rm p}$/R$_{\oplus}$)$^{3}$. 
Only exoplanetary masses with uncertainties less than 25$\%$ were included, leading to a sample of 51 exoplanets. 
Most of the mass measurements for this sample come from radial velocities, with only four exoplanets having masses measured from TTV (\textit{transit timing variations}). For this sample, the median errors in the masses derived via RV and TTV are 15$\%$ and 10$\%$, respectively. We also compared the adopted masses for 18 exoplanets in common with \citet{luque2022}, finding good agreement, with a mean ($\pm$STD) offset $<$``This work - Luque \& Pall\'e"$>$ of +0.03 $\pm$0.12M$_{\oplus}$.

\begin{figure*}
\begin{center}
  \includegraphics[angle=0,width=0.9\linewidth,clip]{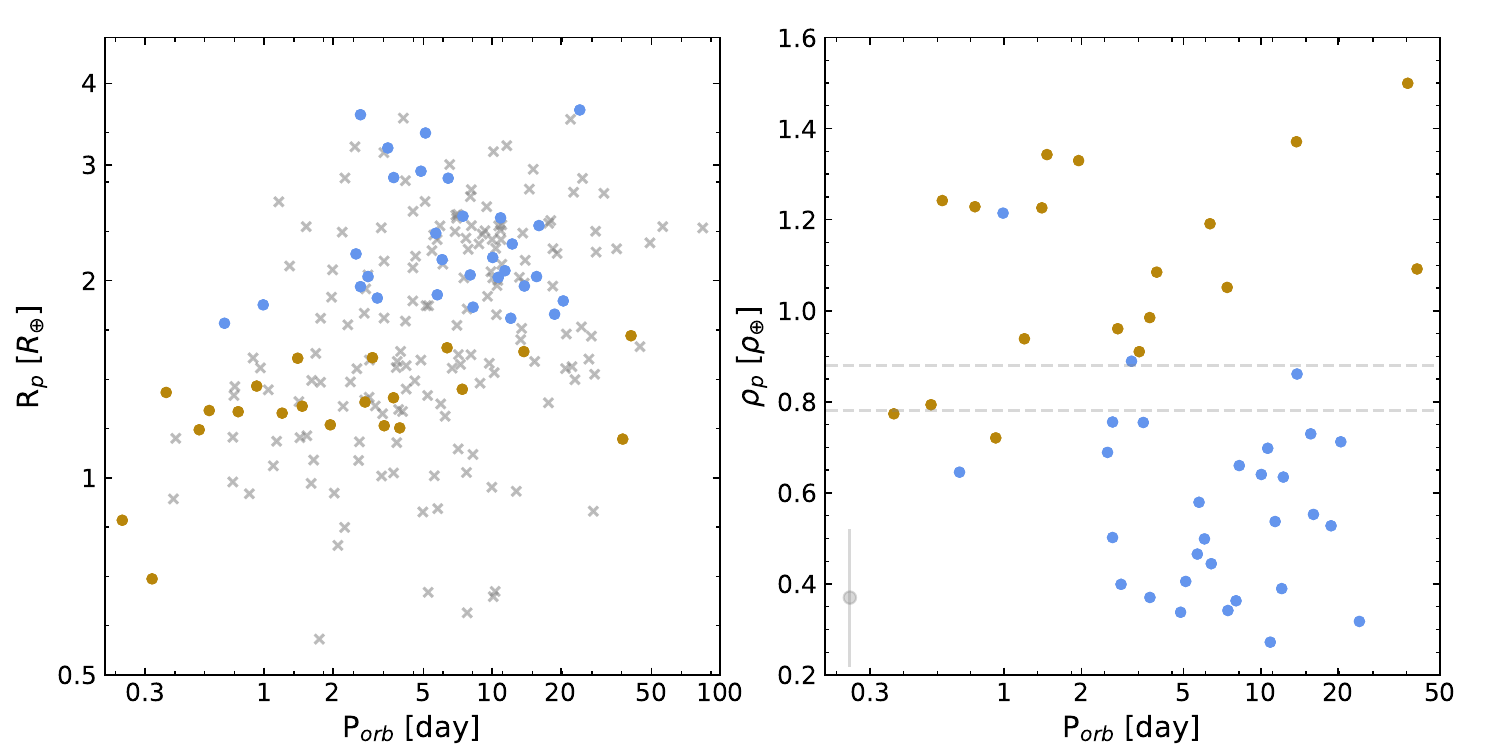}
\caption{Left panel: exoplanetary radii as a function of orbital periods, where sample exoplanets, having masses available, which fall on the ``rocky” side of the gap are colored in gold, while those falling on the sub-Neptune side of the gap are colored in blue. Grey xs are exoplanets in our sample without available mass measurements. Right panel: exoplanetary densities as a function of orbital period following the same color scheme, for exoplanets with $\rho_{\rm p}<$1.6 $\rho_{\oplus}$. Typical uncertainty in the density is shown in the left bottom of the figure.}
\end{center}
\label{dens}
\end{figure*}

Results for radii and densities for our exoplanet sample are presented side by side in the panels of Figure \ref{dens}. In the left panel, we show again the exoplanetary radius as a function of orbital period for the Full sample, but now color those 51 exoplanets with estimated masses in blue (if the exoplanet falls on the sub-Neptune side of the radius gap), or in gold (if the exoplanet falls on the rocky side of the gap). The grey xs in this panel represent the other exoplanets in our sample for which we do not have masses.
The right panel of Figure \ref{dens} shows the results for densities ($<$ 1.6$\rho_{\oplus}$) versus orbital periods, adopting the same color scheme as in the left panel.
The main feature in this panel is the segregation in density between the rocky planets (gold) and sub-Neptunes (blue), where this division between the two families is set by the radius gap from the left panel, with the rocky planets now occupying overall the high-density regime, while the sub-Neptunes fall in the low-density region. 
Another feature in this figure is the appearance of the sub-Neptune desert, but now in density-orbital period space, with the near-absence of the lower-density exoplanets with P$_{\rm orb}$$\lesssim$ 2 days.
The boundary between the rocky and sub-Neptune planets occurs at a density of $\rho_{\rm P}\sim$0.8 - 0.9$\rho_{\oplus}$ (marked by the dashed lines) and appears as a narrow ``density gap'', or separation over this transition region. The nature of the density distribution changes abruptly at this transition gap, with only a small mixture of rocky and sub-Neptune exoplanets mingling together across this gap, which again corresponds to the radius gap projected into density - orbital period space.

Two exoplanets in Figure \ref{dens} deserve some discussion: TOI-1685b and TOI-1634b.  Both have periods less than one day (with P$_{orb}$=0.7 days for TOI-1685b and 1.0 day for TOI-1634b) and are thus members of the ultra-short period (USP) family of exoplanets. Both fall above the radius gap and are thus placed among the sub-Neptune group; these two exoplanets can be found in the panels of Figure \ref{dens} as the blue points with the shortest orbital periods. TOI-1685b has R$_{\rm p}$=1.72 R$_{\oplus}$, as determined by us, and a mass of 3.30 M$_{\oplus}$ (an average of 3.09$^{+0.59}_{-0.58}$ M$_{\oplus}$, \citealt{luque2022}; 3.03$^{+0.33}_{-0.32}$ M$_{\oplus}$, \citealt{burt2024}; 3.78$\pm$0.63 M$_{\oplus}$, \citealt{bluhm2021}), leading to a density of $\rho_{P}$=0.65$\rho_{\oplus}$.  Although our radius and adopted mass are close to those from \citet{bluhm2021}, who derived R$_{\rm p}$=1.70 R$_{\oplus}$ and M$_{\rm p}$=3.78 M$_{\oplus}$, another measurement by \citet{burt2024} found R$_{\rm p}$=1.47 R$_{\oplus}$ and  M$_{\rm p}$=3.03 M$_{\oplus}$ suggesting that in terms of both radius and mean density, TOI-1685b is a rocky world with very little volatile content. The other stand-out among the USP exoplanets in our sample, TOI-1634b, has a radius placing it above the radius gap, with R$_{\rm P}$=1.84 R$_{\oplus}$, while also having a large mass of M$_{\rm p}$= 7.54 M$_{\oplus}$ (an average of 10.14$\pm$0.95 M$_{\oplus}$, \citealt{hirano2021}; 7.57$^{+0.71}_{-0.72}$ M$_{\oplus}$, \citealt{luque2022}; 4.91$^{+0.68}_{-0.70}$ M$_{\oplus}$, \citealt{cloutier2021}) leading to $\rho_{P}$=1.21$\rho_{\oplus}$, placing it squarely, in terms of density, among the rocky planets.

\begin{figure*}
\begin{center}
  \includegraphics[angle=0,width=1\linewidth,clip]{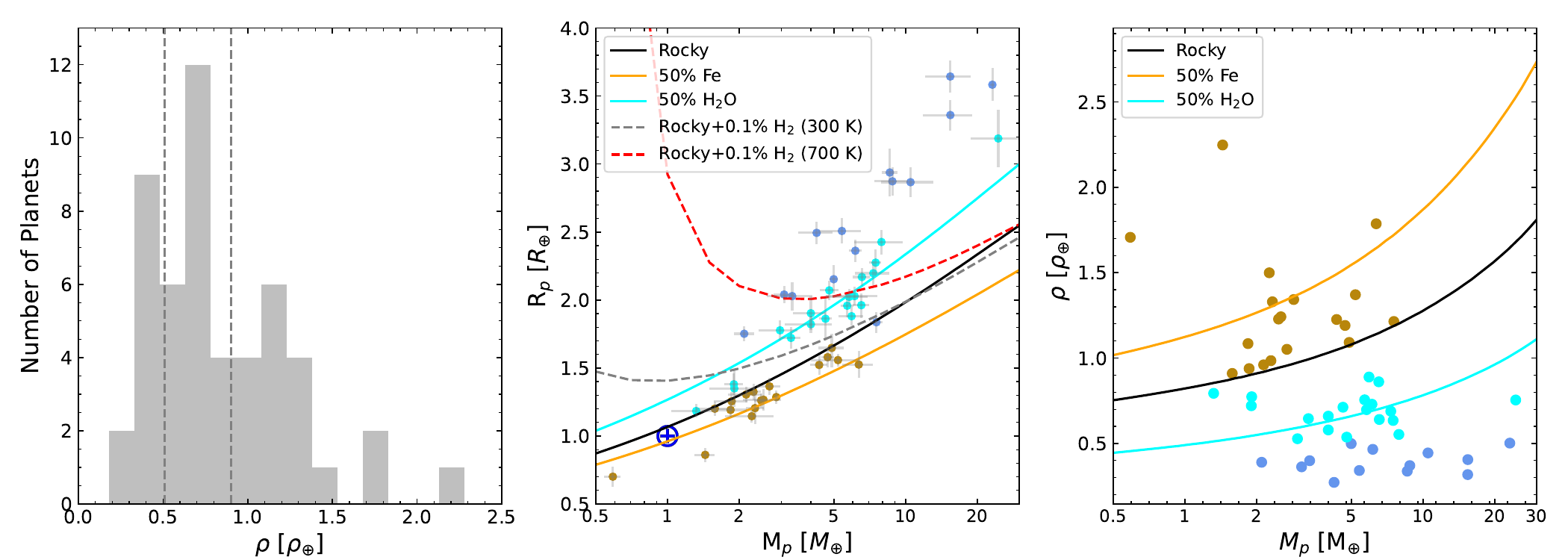}
\caption{Left panel: density histogram of our exoplanetary sample with available masses from the literature. The dashed lines indicate the boundaries for segregating exoplanets into three groups. Middle panel: Derived planetary radii as a function of planetary mass, along with mass-radius relations from \citet{zeng2019} associated with ``rocky'' (black line), ``50$\%$ Fe'' (gold line), and ``50$\%$ H$_{\rm 2}$O'' (cyan line) compositions. We also show gray and red dashed lines representing rocky compositions with 0.1$\%$ H$_{\rm 2}$ atmospheres, for equilibrium temperatures of respectively 300 K and 700 K. Exoplanets that fall between densities between 0.5 - 0.9 $\rho_{\oplus}$ are depicted in cyan. The Earth's position is also shown for comparison. Right panel: the exoplanetary mass - density diagram.}
\end{center}
\label{densother}
\end{figure*}

We take a closer look at the density gap using a histogram of the exoplanet densities shown in the left panel of Figure \ref{densother}.
This distribution shows that, for our data, there is a drop in the number of exoplanets past $\rho_{\rm P}$ $\sim$0.8$\rho_{\oplus}$, which marks the density transition between the rocky planets and sub-Neptunes discussed above.
The rocky planet domain in our sample exhibits a broader distribution, with a peak at $\rho_{\rm P}$ $\sim$1.1$\rho_{\oplus}$, and there are three exoplanets with densities greater than $\rho_{\rm P}$ $\sim$1.5$\rho_{\oplus}$. The sub-Neptune domain in the density histogram displays two peaks, which divide the sub-Neptunes into two possible subgroups, with one group having a distribution around $<\rho_{\rm P}>$=0.4$\rho_{\oplus}$, while the other has $<\rho_{\rm P}>$=0.7$\rho_{\oplus}$, with a decrease in the distribution of densities at $\rho\sim$0.6$\rho_{\oplus}$.

We note that \citet{luque2022} also found two populations in their lower density regime that corresponded to the ``water worlds'' and ``puffy sub-Neptunes''. A comparison of our distribution of exoplanetary densities with that in Figure 3A from \citet{luque2022} finds that the density gap identified here agrees with the location of their gap, when taking into account that their densities are normalized to a model of an Earth-like composition that accounts for gravitational compression from \citet{zeng2019}.

The middle panel of Figure \ref{densother} shows the M$_{\rm p}$ versus R$_{\rm p}$ diagram for the 51 exoplanets plotted by color, according to their different populations as discussed above, with the rocky planets ($\rho_{\rm p}> $0.9$\rho_{\oplus}$) shown as filled gold circles.  The sub-Neptune exoplanets are divided into the lower-density group ($\rho_{\rm p}\leq$0.5$\rho_{\oplus}$), plotted as dark blue symbols, and the intermediate density exoplanets ($\rho_{\rm p}$ between 0.5-0.9$\rho_{\oplus}$), as cyan points. In addition to the exoplanetary radii and masses, five curves representing planetary models from \citet{zeng2019} are shown; the gold curve is a model composed of 50$\%$ Fe, by mass, which is a somewhat larger Fe-fraction relative to the composition of the Earth (which is roughly $\sim$35$\%$ Fe, along with $\sim$30\% O and $\sim$30\% Si+Mg). The black curve from \citet{zeng2019} is their ``Rocky'' model, which has a somewhat larger fraction of Si+Mg relative to Fe when compared to Earth. The cyan curve is a model associated with a 50$\%$ water and 50\% rocky composition. The grey and red dashed lines are models for rocky planets with 0.1$\%$ H$_{\rm 2}$ atmospheres, considering equilibrium temperatures of, respectively, 300 K and 700 K.
The right panel of Figure \ref{densother} shows exoplanetary densities as a function of  masses, plotted as the same colors as in the middle panel, along with some of the models from \citet{zeng2019}.

The positions of the exoplanets shown in cyan (0.7$\rho_{\oplus}$ peak) align quite well with the model for 50$\%$ water (cyan curve) in both the middle and right panels of Figure \ref{densother}, indicating that their mass-radius relations fall at locations that are expected for volatile-rich water worlds (\citealt{luque2022}).
We note, however, that the locations in the mass-radius diagram for some of these planets also fall near the red and gray dashed curves, and therefore are also consistent with rocky planets with modest light-gas atmospheres; this illustrates that there is a degeneracy in the mass-radius diagram versus planetary composition models.
Finally, exoplanets with $\rho_{\rm p}$$>$0.9$\rho_{\oplus}$ (shown as gold symbols), and most of them having radii less than the radius gap (R$_{\rm P}<$1.65 R$_{\oplus}$), for the most part, are found either on, or in between the rocky and 50$\%$ Fe curves, identifying them as 'bona fide' rocky planets.  

\nocite{2023AJ....165..134O,2016ApJS..225....9H,2015ApJ...809...25M,2008ApJ...677.1324T,2024PASA...41...30M,2019ApJS..244...11K,2024ApJ...975L..22W,2023NatAs...7.1317L,2024AJ....168...93J,2024AeA...687A.264B,2024MNRAS.527.5464C,2024AeA...690A.263G,2024AeA...684A..12F,2022AeA...665A.120C,2016ApJS..222...14V,2023AJ....165...84M,2016ApJS..226....7C,2023AeA...675A..52C,2019NatAs...3.1099G,2023AeA...677A..38B,2018Natur.553..477B,2021AeA...645A..41L,2020AJ....160..117R,2021Sci...374.1271L,2012ApJ...750..114F,2016AeA...594A.100B,2023AJ....166..195Q,2023AeA...675A.177G,2020AJ....160..192S,2020AJ....160....3C,2020AJ....159..100S,2018AJ....156...78L,2020AeA...642A..49D,2023AJ....165..167C,2021AJ....162..259Z,2021AJ....162...87B,2017AJ....154..264T,2021AJ....161...13W,2023AJ....165..155T,2022AeA...657A..45S,2021MNRAS.508..195D,2017AJ....154..224R,2022AJ....163..269R,2020AJ....160..259S,2016MNRAS.461.3399P,2018AJ....156..266S,2016MNRAS.463.1780L,2018AJ....156...22Y,2020MNRAS.499.5416C,2009AeA...507..481C,2018MNRAS.480L...1D,2018MNRAS.476L..50D,2018MNRAS.473L.131W,2019MNRAS.487.1865W,2025AJ....169..109T,2024MNRAS.527.5464C,2023AeA...680A..76M,2023AJ....166....9M,2022AJ....163..269R,2022AJ....163...99G,2021AeA...645A..41L,2020AJ....160..259S,2020AJ....160..117R,2019AJ....157..102L,2018MNRAS.480L...1D,2018MNRAS.476L..50D,2015ApJ...801...18M,2023AeA...677A..33B,2024AJ....168..101D,2008ApJ...677.1324T,2024ApJ...975L..22W,2014ApJ...785..126K,2024AeA...688A.216H,2023AeA...675A..52C,2024MNRAS.527.5464C,2021Sci...371.1038T,2023ApJ...955L...3G,2021AeA...653A..41D,2022AeA...665A.120C,2010MNRAS.408.1689S,2024AeA...684A..83M,2023AeA...678A..80P,2023AeA...669A.117L,2021Sci...374.1271L,2016MNRAS.459..789T,2024AeA...685A.147G,2022AeA...664A.199L,2019AeA...628A..39L,2023AeA...675A.177G,2022AeA...665A..91A,2020AJ....160....3C,2020AeA...642A.173N,2024MNRAS.530.4665R,2020AeA...642A.236K,2019AJ....157...97K,2022AJ....164..172D,2021AJ....162...87B,2020AJ....160..192S,2021AeA...645A..41L,2014AeA...572A..73L,2023AJ....165..167C,2020AeA...636A..58A,2021AeA...650A.145C,2021AeA...649A.144S,2019AeA...629A.111C,2022AeA...666A.154L,2014AcA....64..323M,2020AJ....159..120L,2019AJ....158..133H,2018AeA...615A..69D,2016ApJ...823..115D,2023AeA...670A.136K,2018MNRAS.478..460A,2009ApJ...694.1559S}

\section{Conclusions}

This paper investigates the radius gap for small exoplanets orbiting M dwarf stars detected by the Kepler, K2, and TESS missions. Exoplanet radii were derived using transit depth values from the literature, with stellar parameters obtained from high-resolution near-infrared spectra analyzed from the SDSS/APOGEE survey, along with Gaia distances. Synthetic spectra were computed using the LTE radiative transfer code Turbospectrum \citep{plez2012_turbospectrum}, MARCS model atmospheres \citep{gustafsson2008_marcs}, the APOGEE DR17 line list \citep{smith2021}, and adopting the same methodology as in our previous analyses of M dwarf stars (\citealt{wanderley2023}; \citealt{souto2020}). 
The results obtained for our sample of stars with APOGEE spectra were then used to build a calibration of stellar radius as a function of the absolute magnitude, M$_{\rm K_{\rm s}}$, which was then applied to derive radii in an extended sample of M dwarfs to determine radii for a sample of 218 exoplanets. The main conclusions are as follows.

We find that the distribution of exoplanet radii shows a clear peak around 1.2 -- 1.6 R$_{\oplus}$ (``rocky'' peak), which is followed by a drop in the number of exoplanets towards larger radii, delineating a radius gap at R$_{\rm p}$=1.6 -- 2.0 R$_{\oplus}$, plus the presence of a second peak at $\sim$2.0 -- 2.4 R$_{\oplus}$ (``non-rocky'' peak).

The position of the radius gap as a function of orbital period is nearly constant, with very little dependence on orbital periods. Formally, we derive a slope of the power law to be dlog(R$_{\rm p}$)/dlog(P$_{\rm orb}$)=$+$0.01$^{+0.03}_{-0.04}$, with the zero slope being consistent within our uncertainties. This result for M dwarfs is in contrast to the power-law slope for the more massive FGK dwarf hosts of $-$0.11 (e.g., \citealt{martinez2019}).

We also investigated the radius gap as a function of insolation, S$_{\rm p}$, and found a gap roughly between 1.6 -- 2.0 R$_{\oplus}$.
Both the P$_{\rm orb}$ and S$_{\rm p}$ versus R$_{\rm p}$ distributions show a clear sub-Neptune desert, which represents a lack of sub-Neptunes at short orbital periods ($\lesssim$2 days) or high insolation levels, compared to rocky planets. The sub-Neptunes in our sample have S$_{\rm p}\lesssim$120 S$_{\oplus}$, which is much lower than that for FGK stars (at S$_{\rm p}\lesssim$650 S$_{\oplus}$), indicating that the shape of the sub-Neptune desert has a dependence on stellar mass.

Fifty-one exoplanets from our sample had accurate mass measurements in the literature that were used to calculate mean exoplanetary densities. An initial segregation of this sample into two groups, based on whether their respective radii placed them above or below the radius gap, resulted in a near-perfect division in the $\rho_{p}$-P$_{\rm orb}$ plane, where the sub-Neptunes had smaller mean densities and ``rocky'' exoplanets had larger values of $\rho_{p}$, with the transition density gap occurring at $\rho_{p}\sim$0.9$\rho_{\oplus}$.
This indicates that our derived radius gap is efficient in separating the sample into ``rocky'' exoplanets and sub-Neptunes with hydrogen-helium envelopes.

In addition to the gap in the density distribution at 0.9$\rho_{\oplus}$, which separates the rocky planets from the sub-Neptunes (and generally coincides with the gap seen in the exoplanetary radii space), further examination of the exoplanet density distribution revealed not simply two peaks (separating the rocky exoplanets from the sub-Neptunes), but three different peaks, which were used as references to separate our sample based on density intervals: $\rho_{p}$$\leq$0.5$\rho_{\oplus}$, $\rho_{p}$ between 0.5 to 0.9$\rho_{\oplus}$, and $\rho_{p}$$>$0.9$\rho_{\oplus}$ The corresponding mass-radius and mass-density diagrams for the exoplanets within these three groups compared to models by \citet{zeng2019} placed them into ``rocky'' planets, sub-Neptunes, plus water worlds, the latter coinciding with their 50$\%$ H$_{2}$O/50$\%$ Rock model.

We compared the radius gap -- orbital period relations in this work with predictions from models of gas-poor planetary formation \citep{lopez2018}, photoevaporation \citep{lopez2018}, core-powered mass loss \citep{gupta2019}, impact-erosion \citep{wyatt2020}, and pebble accretion \citep{venturini2024}.  
Of these various models, only the pebble accretion models predict near-flat power-law slopes, in general agreement with the results in this study.
The nearly flat slope of the radius gap for M dwarf hosts contrasts with those derived for FGK stars, where photoevaporation and core-powered mass loss are likely mechanisms shaping the radius gap.  
These results suggest that, for M-dwarf systems, inward exoplanetary migration may play a significant role. Exoplanets located further away from the star would be migrating inward and ``filling'' the gap, which would flatten up an originally negative slope originated by photoevaporation / core-powered mass loss, since inward orbital migration is more efficient for small exoplanets around M dwarfs than around solar-type stars.

In addition to the process of inward exoplanetary migration, early findings that the dust disk mass -- stellar mass relation steepens with time, together with evidence that the average solid mass in transiting exoplanets increases towards lower-mass stars, already suggested that solids are redistributed from the outer to the inner disk more efficiently around M dwarfs \citep{pascucci2016}, consistent with enhanced pebble accretion. More recently, the observation that the occurrence rate of transiting sub-Neptunes declines for M dwarfs cooler than ~3,700 K \citep{hardegree2025} is likewise consistent with pebble-accretion models, in which the low frequency of giant planets around low-mass stars drives more efficient redistribution of solids (e.g., \citealt{mulders2021}). These trends carry major implications for the solid and atmospheric composition of planets around M dwarfs, which are beginning to be explored (e.g., \citealt{lin2025}).

In summary, the much flatter power-law relation for R$_{\rm gap}$ as a function of orbital period around exoplanet-hosting M-dwarf stars, when compared to FGK host stars, can be explained by an increased importance of pebble accretion and inward exoplanetary migration relative to photoevaporation and core-powered mass loss in the M-dwarf systems.  Additionally, the sub-Neptune desert extends to significantly lower levels of insolation around M-dwarf hosts. Taken together, these results highlight fundamental differences in the formation of exoplanetary systems in lower-mass stars.

\clearpage
\startlongtable
\begin{deluxetable*}{lccccccccccccccccc}
\small\addtolength{\tabcolsep}{-3.5pt}
\tablenum{1}
\label{fulldata}
\tabletypesize{\scriptsize}
\tablecaption{Stellar and Planetary Data}
\tablewidth{0pt}
\tablehead{
\colhead{Hostname} &
\colhead{d} &
\colhead{T$_{\rm eff}$} &
\colhead{$\log{g}$} &
\colhead{L$_{*}$} &
\colhead{R$_{\rm *,spec}$} &
\colhead{M$_{\rm K_{\rm s}}$} &
\colhead{R$_{*}$} &
\colhead{Planet} &
\colhead{$\Delta$F} &
\colhead{R$_{\rm p}$} &
\colhead{P$_{\rm orb}$} &
\colhead{S$_{\rm p}$} &
\colhead{M$_{\rm p}$} &
\colhead{$\rho_{\rm p}$}\\
\colhead{...} &
\colhead{pc} &
\colhead{K} &
\colhead{...} &
\colhead{10$^{-4}$ L$_{\odot}$} &
\colhead{R$_{\odot}$} &
\colhead{...} &
\colhead{R$_{\odot}$}  &
\colhead{...}  &
\colhead{$\%$}  &
\colhead{R$_{\oplus}$}  &
\colhead{day}  &
\colhead{S$_{\oplus}$}  &
\colhead{M$_{\oplus}$}  &
\colhead{$\rho_{\oplus}$} &
}
\startdata
G 9-40 & 27.82$\pm$ 0.02 & 3423 & 4.89 & 106$\pm$ 2 & 0.29$\pm$0.02 & 6.97 & 0.30$\pm$0.01 & G 9-40 b & 0.34$\pm$0.03\textbf{(1)} & 1.90$\pm$0.12 & 5.75 & 6.60 & 4.00$_{-0.63}^{+0.63}$\textbf{(3)} & 0.58$^{+0.15}_{-0.15}$\\
Kepler-138 & 67.08$\pm$ 0.05 & 3948 & 4.72 & 574$\pm$ 9 & 0.51$\pm$0.03 & 5.37 & 0.52$\pm$0.02 & Kepler-138 c & 0.07$\pm$0.00\textbf{(2)} & 1.56$\pm$0.05 & 13.78 & 6.96 & 5.20$_{-1.20}^{+1.20}$\textbf{(4)} & 1.37$^{+0.34}_{-0.34}$ \\
... & ... & ... & ... & ... &  ... & ... & ... & ... & ... & ... & ... & ...  & ... & ... \\
\enddata
\small
\begin{itemize}[leftmargin=0pt]
    \item[]The T$_{\rm eff}$ and $\log{g}$ uncertainties are respectively $\pm$100 K $\pm$0.20 dex. The complete table is available in electronic format.
    \item[](1) \citet{2020AJ....160..192S,2018AJ....156...22Y}; (2)    \citet{2016ApJS..225....9H,thompson2018}; (3) \citet{2022AeA...666A.154L}; (4) \citet{2018MNRAS.478..460A}
\end{itemize}
\end{deluxetable*}

\acknowledgments

We thank Gijs Mulders for reading the manuscript and providing feedback. We thank the referee for comments that improved the paper. F.W. acknowledges support from fellowship P.C.I, Observatorio Nacional - MCTI.  K.C. and V.S. acknowledge that their work here is supported, in part, by the National Science Foundation through NSF grant No. AST-2009507. D.S. acknowledges support from the Foundation for Research and Technological Innovation Support of the State of Sergipe (FAPITEC/SE) and the National Council for Scientific and Technological Development (CNPq), under grant numbers 794017/2013 and 444372/2024-5.

Funding for the Sloan Digital Sky Survey IV has been provided by the Alfred P. Sloan Foundation, the U.S. Department of Energy Office of Science, and the Participating Institutions. SDSS-IV acknowledges support and resources from the Center for High-Performance Computing at the University of Utah. The SDSS web site is www.sdss.org.
SDSS-IV is managed by the Astrophysical Research consortium for the Participating Institutions of the SDSS Collaboration including the Brazilian Participation Group, the Carnegie Institution for Science, Carnegie Mellon University, the Chilean Participation Group, the French Participation Group, Harvard-Smithsonian Center for Astrophysics, Instituto de Astrof\'isica de Canarias, The Johns Hopkins University, 
Kavli Institute for the Physics and Mathematics of the Universe (IPMU) /  University of Tokyo, Lawrence Berkeley National Laboratory, Leibniz Institut f\"ur Astrophysik Potsdam (AIP),  Max-Planck-Institut f\"ur Astronomie (MPIA Heidelberg), Max-Planck-Institut f\"ur Astrophysik (MPA Garching), Max-Planck-Institut f\"ur Extraterrestrische Physik (MPE), National Astronomical Observatory of China, New Mexico State University, New York University, University of Notre Dame, Observat\'orio Nacional / MCTI, The Ohio State University, Pennsylvania State University, Shanghai Astronomical Observatory, United Kingdom Participation Group,
Universidad Nacional Aut\'onoma de M\'exico, University of Arizona, University of Colorado Boulder, University of Oxford, University of Portsmouth, University of Utah, University of Virginia, University of Washington, University of Wisconsin, Vanderbilt University, and Yale University.

\facility {Sloan}

\software{Matplotlib (\citealt{Hunter2007_matplotlib}), Numpy (\citealt{harris2020_numpy}), Turbospectrum (\citealt{plez2012_turbospectrum})}

\bibliographystyle{yahapj}
\bibliography{references.bib}

\end{document}